\documentclass[doc,natbib]{apa6}
\usepackage{amsmath}
\usepackage{graphicx}

\usepackage{tikz}
\usetikzlibrary{bayesnet}
\usetikzlibrary{arrows}

\usepackage[]{hyperref}
\hypersetup{
   pdfauthor={AUTHORS},
   pdftitle={TITLE},
   pdfkeywords={KEYWORDS},
   breaklinks,
   linkcolor={blue},
   urlcolor={blue},
   anchorcolor={blue},
   citecolor={blue}
}

\title{A hierarchical Bayesian model for measuring individual-level and group-level numerical representations}
\shorttitle{Bayesian model of numerical representations}
\author{Thomas J. Faulkenberry}
\affiliation{Department of Psychological Sciences, Tarleton State
  University, USA}

\authornote{Please address correspondence to Thomas J. Faulkenberry,
  Department of Psychological Sciences, Box T-0820, Tarleton State University,
  Stephenville, TX 76402. Email: faulkenberry@tarleton.edu. All raw
  data and R scripts are available to download at
  http://github.com/tomfaulkenberry/bayesRegression/.  The author would like to thank Lisa Fazio for providing data for Model 2.}

\abstract{
  A popular method for indexing numerical representations is to compute an individual estimate of a response time effect, such as the SNARC effect or the numerical distance effect.  Classically, this is done by estimating individual linear regression slopes and then either pooling the slopes to obtain a group-level slope estimate, or using the individual slopes as predictors of other phenomena.  In this paper, I develop a hierarchical Bayesian model for simultaneously estimating group-level and individual-level slope parameters. I show examples of using this modeling framework to assess two common effects in numerical cognition: the SNARC effect and the numerical distance effect. Finally, I demonstrate that the Bayesian approach can result in better measurement fidelity than the classical approach, especially with small samples.\\

 }
\begin{document}
\maketitle

The purpose of this paper is to introduce a hierarchical Bayesian model for measuring two types of mental representations that are often assessed in numerical cognition: the SNARC effect \citep{dehaene1990} and the numerical distance effect \citep{moyer1967}. The SNARC effect (Spatial-Numerical Associations of Response Codes) refers to the finding that response times (RTs) for small numerical stimuli are faster with the left hand, whereas RTs for large numbers are faster with the right hand. The numerical distance effect (NDE) refers to the finding that RTs decrease as the numerical distance between stimulus numbers increases. Both effects are thought to reflect particular aspects of mental representation of number.  As the SNARC effect occurs even when number magnitude is not salient to a given task (e.g., asking whether a given number is odd or even), the SNARC effect is often taken as an index of automatic processing of number magnitude \citep{wood2008}.  Similarly, the NDE is thought to represent a noisy, analog representation of number \citep{dehaene1992}, and thus is purported to occur because as numerical distance increases, the amount of representational overlap (and hence response competition) decreases, resulting in faster RTs \citep{gevers2006}.  

As indices of numerical representation, both the SNARC effect and the NDE have been used to study individual differences in number representation \citep{holloway2009,viarouge2014}.  For example, \citet{viarouge2014} showed that participants with stronger SNARC effects tended to exhibit slower mental rotation processing speed.  Similarly, \citet{pinhas2014} found that the SNARC effect was positively associated with spatial operational momentum.  The NDE has been used in a similar manner.  \citet{holloway2009} found that children with a larger NDE tended to score lower on standardized assessments of mathematical fluency and calculation skill.  \citet{fazio2014} showed a similar pattern, namely that the NDE was negatively correlated with mathematics achivement, but not reading achievement.  As such, it is clear that accurate measurement of the SNARC effect and NDE is a desirable objective for a wide range of research paradigms within mathematical cognition.

The classical method for measuring the SNARC effect and NDE for individuals is based on a procedure known as regression coefficient analysis \citep[RCA;][]{lorch1990}. In this procedure, a dependent measure such as RT is regressed against a predictor for each individual participant.  The estimated slope and intercept for each individual is then recorded, at which point any number of analyses may be performed.  One of the first examples of this procedure in numerical cognition comes from \citet{fias1996}, who adapted this method to measure the SNARC effect.  Participants' left-hand and right-hand RTs for each number stimulus (e.g., 1, 2, 8, 9) were aggregated via the median.  Then, the difference between right- and left-hand RTs ($dRT$) was computed for each number.  The SNARC effect was then presented as a negative correlation between number and dRT.  That is, for small numbers (e.g., 1,2), the left hand is faster, so $dRT>0$.  For large numbers, the right hand is faster, so $dRT<0$.  This negative correlation is captured by a linear regression, which is used to calculate the slope on $dRT$ for each participant.  Analyses of the NDE are similar, but instead of using $dRT$, one typically regresses $RT$ against a predictor such as numerical distance \citep[e.g.,][]{holloway2009} or ratio \citep[e.g.,][]{fazio2014}. As with the SNARC effect, the NDE typically presents as a negative slope. These slopes can then be submitted to a variety of analyses, including group-level tests (e.g., does the SNARC/NDE effect vary by group?) or individual-level correlations (i.e., is the SNARC/NDE effect correlated with mathematical performance?).  

In the following sections, I describe a hierarchical Bayesian approach to measuring the SNARC effect and the NDE. Certainly, a complete treatment of Bayesian methods is beyond the scope of this paper, but the interested reader is advised consult the excellent books by \citet{gelman2013}, \citet{mcelreath2015}, or \citet{kruschke2015} for more details. However, I will briefy outline some advantages to using such an approach to measuring numerical representations. One such advantage is that in a hierarchical model, variability from participants and items is modeled simultaneously, resulting in in both participant-level and group-level parameter estimates.  As such, the hierarchical model allows the researcher to avoid aggregating single estimates across individuals, which can be problematic \citep{heathcote2000,haider2002,pratte2010}. Another advantage is both philosophical and pragmatic. Bayesian modeling provides a principled means to combine prior knowledge with new data via Bayes' theorem.  The result is a quantification of \emph{posterior} belief in terms of a probability distribution that reflects uncertainty after seeing data. From this distribution, one can estimate a variety of summary statistics, including the posterior mode (i.e., the parameter estimate with highest density), and Bayesian credible intervals, which give a range of values containing a particular probability mass.\footnote{Bayesian credible intervals are not the same thing as frequentist \emph{confidence} intervals, which are defined as a range of parameters that would contain a population parameter in some specified percentage of a large number of repeated samples.  While people often misinterpret confidence intervals using probability statements \citep{hoekstra2014}, Bayesian credible intervals \emph{can} be directly interpreted as probability statements} One particular type of credible interval used in a Bayesian context is the highest posterior density interval (HPDI), which is the \emph{narrowest} interval containing a given probability mass.

Generally, the models will be constructed as follows.  As with the classical method described above, the goal is to estimate a slope on $dRT$ (for the SNARC effect) or $RT$ (for the NDE).  However, the proposed models are hierarchical, which means that individual-level slopes are drawn from a group-level distribution that specifies how the slopes are distributed in the population.  Thus, individual-level and group-level estimates are modeled simuntaneously, which improves upon the noisy slope estimates usually obtained from classical regression with small samples \citep{gelmanHill2013}.  Further, since the model is Bayesian, one can specify group-level prior distributions that will moderate extreme individual-level parameter estimates (i.e., shrinkage), which will result in increased measurement accuracy for our parameter estimates.

\section{Model 1: estimating the SNARC effect}

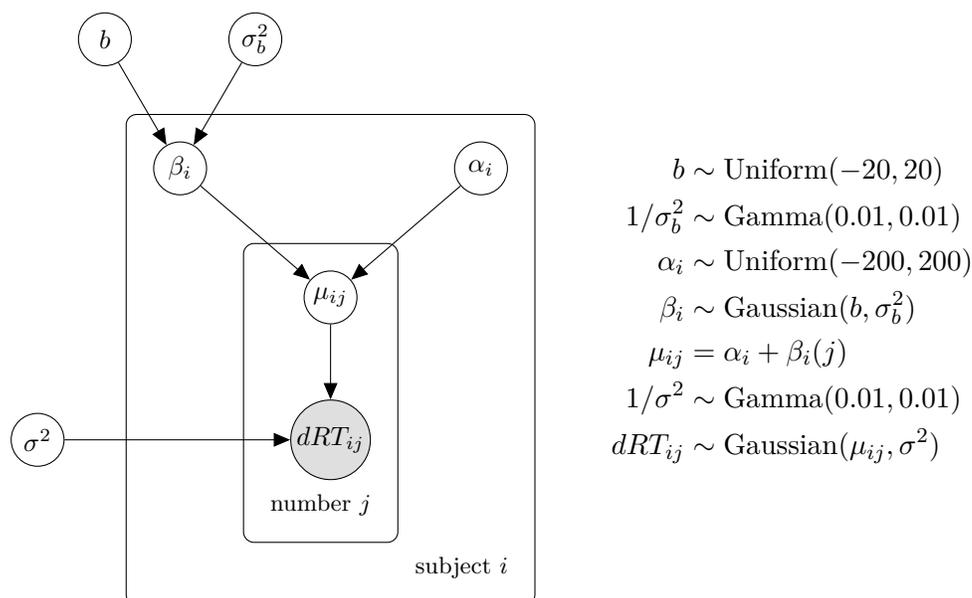
\begin{figure}
  \centering
\begin{tikzpicture}

  \node[obs] (dRT) {$dRT_{ij}$};%
  \node[latent, above=of dRT] (mu) {$\mu_{ij}$}; %
  \node[latent, above=of mu, xshift=-2cm] (beta) {$\beta_i$};%
  \node[latent, above=of mu, xshift=2cm] (alpha) {$\alpha_i$};%

  \node[latent, above=of beta, xshift=-1cm] (b) {$b$};%
  \node[latent, above=of beta, xshift=1cm] (sigma_b) {$\sigma_b^2$};%

  \node[latent, left=of dRT, xshift=-2cm] (sigma) {$\sigma^2$};%
     
     \edge {sigma,mu} {dRT}
     \edge {beta,alpha} {mu}
     \edge {b,sigma_b} {beta}

     \plate[inner sep=10pt] {plate1} {(dRT)(mu)} {number $j$}; %
     \plate[inner sep=10pt] {plate2} {(plate1)(beta)(alpha)} {subject $i$}; %

     \node[text width=6cm, anchor=west, right] at (3,2)
     {
       \begin{align*}
         b & \sim \text{Uniform}(-20,20)\\
         1/\sigma_b^2 & \sim \text{Gamma}(0.01,0.01)\\
         \alpha_i & \sim \text{Uniform}(-200,200)\\
         \beta_i & \sim \text{Gaussian}(b,\sigma_b^2)\\
         \mu_{ij} &= \alpha_i + \beta_i(j)\\
         1/\sigma^2 &\sim \text{Gamma}(0.01, 0.01)\\
         dRT_{ij} & \sim \text{Gaussian}(\mu_{ij},\sigma^2)
       \end{align*}
};
 \end{tikzpicture}

  \caption{Graphical model of the hierarchical Bayesian linear regression model for the SNARC effect. Following the convention of \citet{lee2014}, nodes represent variables of interest (observed=shaded, latent=unshaded), with dependencies represented via the graph structure.}
  \label{fig:model}
\end{figure}

The model is a hierarchical Bayesian linear regression model, where $dRT$ is predicted by stimulus number.  A graphical representation of the model can be seen in Figure \ref{fig:model}.  Formally, we define

\begin{equation}\label{eq:model}
  dRT_{ij} = \alpha_i+\beta_i(j) + \varepsilon_{ij}
\end{equation}

\noindent
where $i =$ subject number and $j=$ stimulus number.  The residuals $\varepsilon_{ij}$ are assumed to be normally distributed with mean 0 and precision $1/\sigma^2$.  Thus, we can express the likelihood for the data as

\begin{equation}\label{eq:likelihood}
 dRT_{ij} \sim \text{Normal}(\mu_{ij}, 1/\sigma^2)
\end{equation}

\noindent
where $\mu_{ij}=\alpha_i+\beta_i(j)$.  The prior for each individual-level intercept $\alpha_i$ is set to be uniformly distributed between -200 and 200.  The hierarchical structure is instantiated on slope.  First, I set the prior for each individual-level slope $\beta_i$ to be normally distributed with two hyperparameters: mean $b$ and precision $1/\sigma_b^2$.  In turn, this requires priors on the hyperparameters -- the prior for $b$ is uniformly distributed between -20 and 20.  Note that this range is based on inspection of the slopes obtained in \citet{viarouge2014}.  The prior for group-level slop precision $1/\sigma_b^2$ (as well as group-level residual precision $1/\sigma^2$) is set as a Gamma distribution with both shape and scale equal to 0.01. 

\subsection{Fitting the model}

\subsubsection{Data}
I fit the model to data collected from 35 participants in a number parity task.  The numbers 1, 2, 8, and 9 were presented in the center of a computer screen, after which participants were asked to quickly indicate via a button press whether the number was even or odd.  The procedure mirrored that of Experiment 1 of \citet{pinhas2014}.  Participants completed 112 trials of the task under two counterbalanced response rules (either even=left or even=right).  This resulted in a collection of 3,920 trials.  We removed 259 error trials and an additional 12 trials for which RT exceeded 3 seconds (a total of 6.9\% of trials).  The remaining 3,649 trials were collapsed into $35 \times 8 \times 2$ cells by computing median $RT$ for each of the conditions defined by crossing the factors of subject, stimulus number, and response hand.  Then, $dRT$ was computed for each combination of subject and number by subtracting left-hand RT from right-hand RT.

\subsubsection{Results}
The regression model parameters were estimated using R \citep{r} and JAGS \citep{jags}.  Posterior sampling consisted of 3 MCMC chains, each containing 100,000 draws.  The first 5000 draws of each chain were discarded as ``burn-in'' samples, leaving 285,000 samples remaining.  These remaining samples were thinned by a factor of 10, leaving a final sample of 28,500 posterior draws for each parameter.  Visual inspection of trace plots indicated that all chains converged appropriately.  Additionally, the Gelman-Rubin statistic $\hat{R}=1.001$ for each parameter, indicating that the Markov chains for each parameter converged to the appropriate stationary distribution \citep{gelman1992,gelman2013}.

Since the model is hierarchical, I was able to estimate posterior distributions for each $\alpha_i$ and $\beta_i$ (i.e., each participant's intercept and slope, respectively). Further, I estimated the posterior distribution of $b$, which is the group level mean slope. The flexibility of this model allows one to ask many questions about the SNARC effect, both at the group level and the individual level.  To illustrate, I will investigate whether there was an overall SNARC effect for the group \citep[c.f.,][]{fias1996}. This can be answered by looking at the posterior distribution of the group-level $b$.

\begin{figure}
  \centering
  \includegraphics[width=0.9\linewidth]{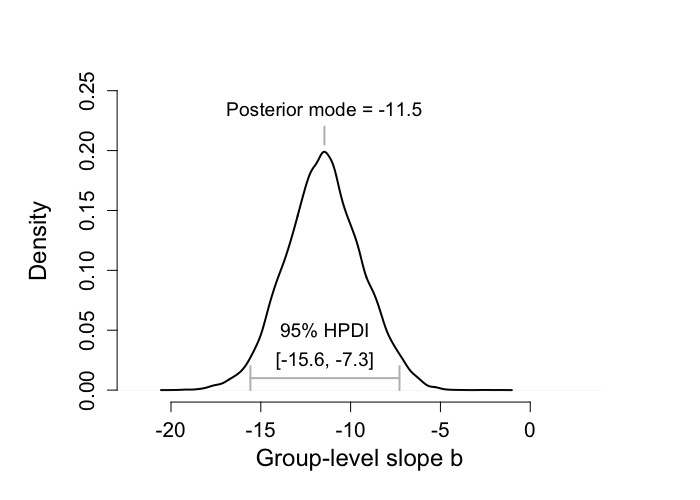}
  \caption{Posterior distribution of the group-level slope $b$ for Model 1, indicating a substantial SNARC effect across the group.}
  \label{fig:model1-posterior}
\end{figure}

The posterior distribution of the group-level slope $b$ is depicted in Figure \ref{fig:model1-posterior}.  As can be seen in the figure, the mass of the distribution is centered over the posterior mode $b=-11.5$.  Further, the 95\% HPDI for the slope $b$ is $[-15.6, -7.3]$.  Finally, since the posterior distribution is a \emph{probability distribution}, we can compute the probability that $b<0$ (that is, the probability that there is a non-zero SNARC effect).  This probability is greater than 0.999.  Other Bayesian tools can be applied to this model, such as computing a Bayes factor comparing a null-SNARC model to our obtained model via the Savage-Dickey density ratio \citep{wagenmakers2010}. Briefly, this method amounts to computing the density of the value $b=0$ in the posterior distribution divided by the density of $b=0$ in the prior (i.e., $\text{Uniform}(-20,20)$).  Doing this resulted in a Bayes factor of approximately 280,000 to 1 in favor of the alternative hypothesis, indicating overwhelming support for an overall SNARC effect.

\section{Model 2: estimating the numerical distance effect}

\begin{figure}
  \centering

\begin{tikzpicture}

  \node[obs] (RT) {$RT_{ij}$};%
  \node[latent, above=of RT] (mu) {$\mu_{ij}$}; %
  \node[latent, above=of mu, xshift=-2cm] (beta) {$\beta_i$};%
  \node[latent, above=of mu, xshift=2cm] (alpha) {$\alpha_i$};%

  \node[latent, above=of beta, xshift=-1cm] (b) {$b$};%
  \node[latent, above=of beta, xshift=1cm] (sigma_b) {$\sigma_b^2$};%

  \node[latent, left=of dRT, xshift=-2cm] (sigma) {$\sigma^2$};%
     
     \edge {sigma,mu} {dRT}
     \edge {beta,alpha} {mu}
     \edge {b,sigma_b} {beta}

     \plate[inner sep=10pt] {plate1} {(dRT)(mu)} {number $j$}; %
     \plate[inner sep=10pt] {plate2} {(plate1)(beta)(alpha)} {subject $i$}; %

     \node[text width=6cm, anchor=west, right] at (3,2)
     {
       \begin{align*}
         b & \sim \text{Uniform}(-100,100)\\
         1/\sigma_b^2 & \sim \text{Gamma}(0.01,0.01)\\
         \alpha_i & \sim \text{Uniform}(0,2000)\\
         \beta_i & \sim \text{Gaussian}(b,\sigma_b^2)\\
         \mu_{ij} &= \alpha_i + \beta_i(j)\\
         1/\sigma^2 &\sim \text{Gamma}(0.01, 0.01)\\
         dRT_{ij} & \sim \text{Gaussian}(\mu_{ij},\sigma^2)
       \end{align*}
};
 \end{tikzpicture}

  \caption{Graphical model of the hierarchical Bayesian linear regression model for the numerical distance effect. Following the convention of \citet{lee2014}, nodes represent variables of interest (observed=shaded, latent=unshaded), with dependencies represented via the graph structure.}
  \label{fig:model2}
\end{figure}
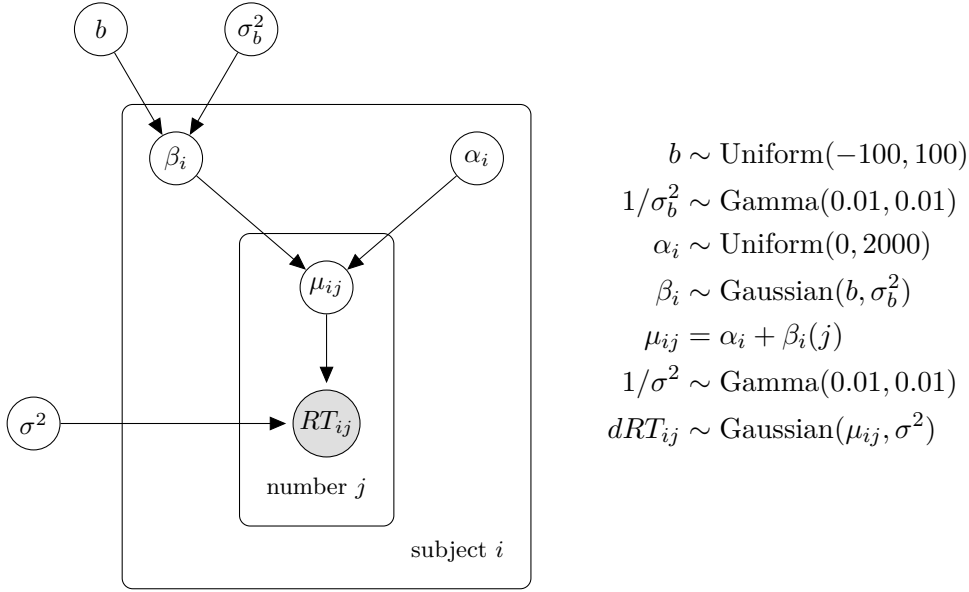

As with Model 1, this model is also a hierarchical Bayesian linear regression model, where $RT$ is predicted by numerical distance. For this specific instantiation of the model, I will index numerical distance via the \emph{ratio} between compared numbers.  For example, \citet{fazio2014} used four ratio bins as predictors of RT.  A graphical representation of the model can be seen in Figure \ref{fig:model2}.  Formally, we define

\begin{equation}\label{eq:model2}
  RT_{ij} = \alpha_i+\beta_i(j) + \varepsilon_{ij}
\end{equation}

\noindent
where $i =$ participant number and $j=1,\dots,4$ equals ratio bin number.  The residuals $\varepsilon_{ij}$ are again assumed to be normally distributed with mean 0 and precision $1/\sigma^2$.  The likelihood for the data is

\begin{equation}\label{eq:likelihood}
 RT_{ij} \sim \text{Normal}(\mu_{ij}, 1/\sigma^2)
\end{equation}

\noindent
where $\mu_{ij}=\alpha_i+\beta_i(j)$.  The prior for each individual-level intercept $\alpha_i$ is set to be uniformly distributed between 0 and 2000. The prior for each individual-level slope $\beta_i$ to be normally distributed with two hyperparameters: mean $b$ (with uniform prior between -100 and 100) and precision $1/\sigma_b^2$ (with a $\text{Gamma}(.01,.01)$ prior).  Finally, the prior for the group-level residual precision is $\text{Gamma}(.01,.01)$.

\subsection{Fitting the model}

\subsubsection{Data}
I fit the model to the RT data from \citet{fazio2014}.  Fifty-five 5th graders completed a symbolic number comparison task, in which they were asked to choose which of two Arabic numerals was larger. Each child completed 40 trials using stimulus numbers ranging from 5 to 21.  Of the 40 trials, 10 came from each of 4 ratio bins: 1.15 - 1.28, 1.28 - 1.43, 1.48 - 1.65, and 2.46 - 2.71. The ratio for each stimulus pair was defined as the quotient obtained when dividing the larger number by the smaller number.  In all, participants completed 2,200 trials.  I removed 95 error trials and 6 additional trials for which RT exceeded 5 seconds (a total of 4.6\% of trials). The remaining 2,099 trials were collapsed into $55 \times 4$ cells by computing median $RT$ for each of the conditions defined by crossing the factors of subject and ratio bin.

\subsubsection{Results}
The regression model was fit using the same procedure as in Model 1.  As before, visual inspection of trace plots indicated that all chains converged appropriately, with $\hat{R}=0.001$ for all parameters.

\begin{figure}
  \centering
  \includegraphics[width=0.9\linewidth]{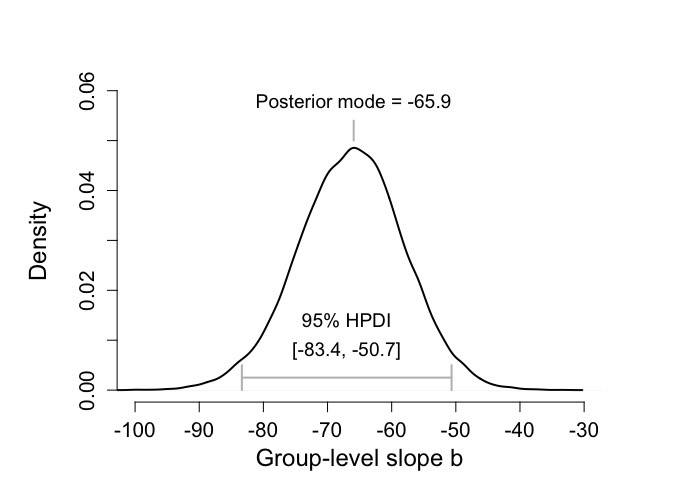}
  \caption{Posterior distribution of the group-level slope $b$ for Model 2, indicating a substantial numerical distance effect across the group.}
  \label{fig:model2-posterior}
\end{figure}

The posterior distribution of the group-level slope $b$ is depicted in Figure \ref{fig:model2-posterior}.  As can be seen in the figure, the mass of the distribution is centered over the posterior mode $b=-65.9$, with 95\% HPDI for the slope $b$ equal to $[-83.4, -50.7]$.  In addition to estimating the group-level slope, we can test the existence of the numerical distance effect via a Bayes factor, which as before I computed using the Savage-Dickey density ratio.  This Bayes factor was approximately 1.7 million to 1 in favor of the alternative hypothesis, indicating overwhelming support for an overall numerical distance effect.

\section{Comparing the model to the classical approach}
In this section, I will demonstrate that the hierarchical Bayesian model developed in this paper can provide better measurement fidelity for the group-level regression slopes that are desired when assessing group-level SNARC effects or NDEs, particularly for small samples.  As mentioned earlier, the reason for this increased accuracy is because of a property that is unique to Bayesian inference -- namely the property of \emph{shrinkage}.  When one specifies a group-level prior distribution for the slope (e.g., the $\text{Uniform}(-20,20)$ distribution that I used in Model 1), this prior is then combined with the data likelihood via Bayes theorem by multiplication.  In our case, extreme individual slope estimates (those beyond -20 and 20) vanish by virtue of being multiplied by the prior probability of obtaining those estimates (i.e., probability = 0).  The resulting posterior distribution is then \emph{shrunk} away from these parameter estimates.

To demonstrate this, I conducted a simulation.  I randomly generated $dRT$ values for a small samples of $n=15$ simulated participants as follows.  First, I generated 15 random slopes $b_i$, where $b_i\sim \text{Gaussian}(-10,1)$.  That is, I assumed that individual slopes are drawn from a normal distribution centered at -10, with standard deviation 1.  Similarly, I randomly generated 15 random intercepts $a_i$, where $a_i \sim \text{Uniform}(-200,200)$.  Then, I generated 

\[
  dRT_{ij} = a_i + b_i(j) + \varepsilon_{ij}
\]

\noindent
where $i=1,\dots,15$, $j=1,2,8,9$, and $\varepsilon_{ij} \sim \text{Gaussian}(0,100)$.  Then, I computed a classical linear regression for each participant $i=1,\dots,15$, recording each estimated slope $\hat{b}_i$.  Finally, I fit the hierarchical Bayesian model (Model 1) to the overall set of data to compute the posterior distribution of $b$.

\begin{figure}
  \centering
  \includegraphics[width=0.9\linewidth]{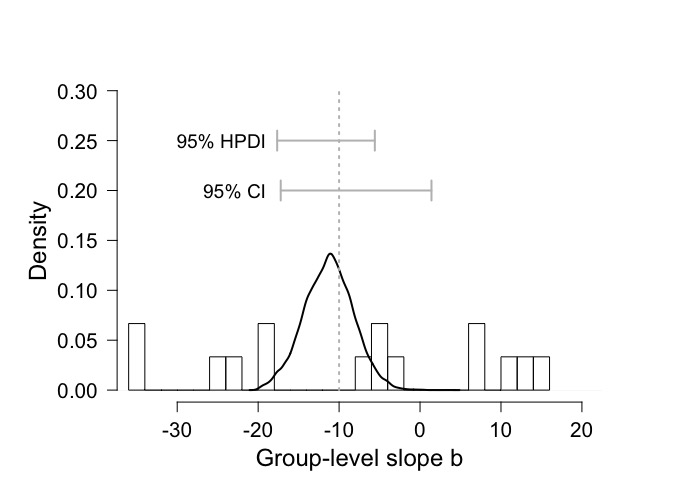}
  \caption{Slope estimates from a simulated data set of $n=15$ participants.  The solid line represents the posterior density of $b$, the group-level slope estimate for the SNARC effect.  The histogram represents the slope estimates obtained from separate classical linear regressions for each individual.}
  \label{fig:simulation}
\end{figure}

The results can be seen in Figure \ref{fig:simulation}.  Notice that while both the frequentist 95\% confidence interval and the Bayesian 95\% highest posterior density interval contain the population mean $b=-10$, the Bayesian estimate is much less variable.  This is because compared to the classical linear regression method (e.g., Lorch \& Myers, 1990), the extreme parameter estimates are shrunk toward the posterior mean.  The same story is repeated with hypothesis testing, as well.  Indeed, performing a $t$-test on the individual regression slopes \citep[as in][]{fias1996} results in a non-significant SNARC effect, $t(14)=-1.82$, $p=0.09$, which results in a Type II error.  However, a Bayesian hypothesis test (the Savage-Dickey method) yields a Bayes factor that favors the SNARC effect by a factor of 75 to 1. 

\section{Discussion}

The purpose of this paper was to introduce a hierarchical Bayesian linear regression framework for measuring group-level and individual-level numerical representations. The models were then applied to two common markers of numerical representations: the SNARC effect, which indexes spatial-numerical associations, and the numerical distance effect (NDE), which indexes representations of numerical magnitude. The models developed here represent a straightforward Bayesian extension of the classical linear regression approach to measuring these effects that was introduced by \citet{lorch1990} and first applied to the SNARC effect by \citet{fias1996}.

Both models were built from a hierarchical Bayesian framework, which is advantageous for several reasons.  First, the models assume that the regression slopes (the primary object of interest in these analyses) are drawn from a group-level distribution. This hierarchical definition allows both group-level and individual-level slope estimates to be modeled simultaneously from the data. As such, we can answer questions at either the group-level (i.e., is there a SNARC effect?) or participant-level (i.e., are the slopes associated with ohter measures, such as math achievement?).  For example, one might use this framework as a stepping stone to a more complex model, where the group-level slopes $b$ are hypothesized to differ by some independent variable. A concrete instance of this comes from studies in which the authors found group differences in the SNARC effect \citep[e.g.,][]{fischer2010,cipora2015}.  One could perform similar studies by building an additional linear model on these group-level parameters, with priors appropriate to such effects \citep[e.g., a Cauchy prior, as in][]{rouder2009}. Importantly, since the inference would be done in a Bayesian framework, it is possible to measure evidence for null effects too \citep{wagenmakers2007}, so the model could be used to test for \emph{invariances} as well as differences. 

Another advantage to using a hierarchical Bayesian model for numerical representations is that such models tend to have better measurement fidelity. Indeed, one of the advantages of the Bayesian framework in general is the notion of shrinkage, where extreme parameter estimates are ``shrunk'' toward the mean by virtue of the prior \citep{gelman2013}. This property is particularly salient for small samples, where classical frequentist methods tend to perform poorly.  I demonstrated exactly this phenomenon in the simulation above: because of some extreme individual-level slope estimates, the frequentist confidence interval was quite wide, and as a result, we could not detect the SNARC effect.  However, the Bayesian highest posterior density interval provided a much more accurate estimate of the true population slope.  Critically, in this simulation, only the Bayesian method produced the correct inference.

A final advantage to this modeling framework that I will mention is that all assumptions of the model are made explicit. This may be a new approach to some, especially those who are accustomed to classical inferential software packages for which the statistical assumptions are kept ``under the hood.''  However, I think the approach presented in this paper can be very useful to a wide variety of problems in cognitive psychology, as the researcher can take the models presented here and modify them for any desired context.  Indeed, the prior knowledge of a given field can be easily and coherently integrated into the model without too much work.

In summary, the models developed in this paper will provide a flexible tool that can be used to estimate group-level and individual-level numerical representations quickly and coherently.  Further, the methods developed are not specific to numerical cognition, so application to a wide variety of problems should not be too difficult. Regardless of the application, the hierarchical Bayesian linear regression framework presented here provides a powerful, coherent, and accurate measurement model for applied work in cognitive psychology.

\bibliography{references}
\bibliographystyle{apalike}

\end{document}